# Mirrored Transformation Optics

JUNKE LIAO [1], PENGFEI ZHAO [1], ZHIBING ZHANG [1], WEN XIAO [1, 2, 3], AND HUANYANG CHEN [1, 2, 4]

[1]Department of Physics, Xiamen University, Xiamen, Fujian, 361005, China
[2]Jiujiang Research Institute of Xiamen University, Jiujiang, Jiangxi, 332000, China
[3]e-mail: 15162148216@163.com
[4]e-mail: kenyon@xmu.edu.cn

**Abstract**
A mirrored transformation optics (MTO) approach is presented to overcome the material mismatch in transformation optics. It makes good use of the reflection behavior and introduce a mirrored medium to offset the phase discontinuities. Using this approach, a high-performance planar focusing lens of transmission-type is designed, which has large concentration ratio than other focusing lens obtained by generalized Snell law. The MTO will not change any functionality of the original lens and promising potential applications in imaging and light energy harvesting.

**Introduction**

Transformation optics (TO) establishes the equivalence between medium and geometry, providing powerful strategies to design electromagnetic functional devices. TO became flourish since 2006, where the two transformation methods, i.e., coordinate transformation and conformal mapping, were successfully used to achieve invisibility cloaks [1-3]. Numerous remarkable devices have been designed in the vein of TO, for instance, field concentrators [4-5], rotators [6,7], illusions [8-9], to name a few. Apart from these, TO has been widely utilized to adapt the performance of optical devices, including waveguides [10-15], lenses [16-18], antennas [19-21]. As a generic solution, TO's concept has been extended to other waves, such as acoustic [22, 23], elastic [24, 25] and liquid surface waves [26, 27].

The core of TO is the transformation formula. Ingenious transformation and achievable transformed media are both important. Material mismatch will occur when the virtual space involved in transformation optics is not full space. Optical mismatch generally has three kinds: impedance mismatch, refractive index mismatch, and phase discontinuity. Since the principle of transformation naturally results in $\boldsymbol{\varepsilon} = \boldsymbol{\mu}$, the impedance is always matched. When polarization is considered, the impedance mismatch may arise [28, 29]. Refractive index mismatch is more common, it can cause unwanted reflection and sacrifice some performance of the device. Several methods can help improve this problem [30, 31]. Sometimes the refractive index mismatch caused by transformation optics is also useful and can be applied to the design of TO micro-cavities with high quality factor and robust unidirectional emission [32-34]. However, the phase discontinuity in TO has not been paid enough attention. No matter the coordinate transformation or conformal mapping, as far as the space shapes are changed during the transformation, phase discontinuity will inevitably occur [35, 36]. Many transformations in TO are not shape-invariant. Phase control is important in many optical applications, such as imaging and unidirectional radiators. Although generalized Snell's law [37, 38] can predict wave propagation under phase discontinuities, there is currently no straightforward approach to completely circumvent this issue in TO. In this paper, we propose a general method to solve the problem of phase discontinuity in traditional TO, which is called Mirrored Transformation Optics (MTO). We first briefly introduce the problems caused by the phase discontinuity during the TO, then we use the MTO to solve it, and finally, we design a highly efficient transmission-type planar focusing lens based on this method. The lens is expected to be used in solar energy harvesting and antennas.

**Results**

The phase discontinuity can be clearly depicted in a compressed transformation shown in Fig. 1. From the TO perspective, as illustrated in Fig. 1 (a, b), when an air trapezoid is compressed into a planar slab, the parallel ray will deflect the same angle as the $\theta$ in trapezoid [39], i.e., $\theta_t = \theta$. However, it is not true in the calculation of Hamiltonian optics [40] as illustrated by the red arrow in Fig. 1(b). We draw the mathematical comparison between these two methods in Fig. 1(e). In the small-angle $\theta$ transformation, the deflection angles in TO perspective $\theta_t$ and Hamiltonian optics $\theta_{tH}$ are nearly indistinguishable. As $\theta$ gradually increases, the $\theta_t$ and $\theta_{tH}$ diverge significantly due to the distinct phase discontinuity at $x = d$. The simulated wave patterns under small and large angle transformations are presented in Figs. 1(c) and (d), respectively. It can be seen that at a large $\theta$ there is obvious wavefront deformation in the plate and the $\theta_t$ is far away from the propagation of

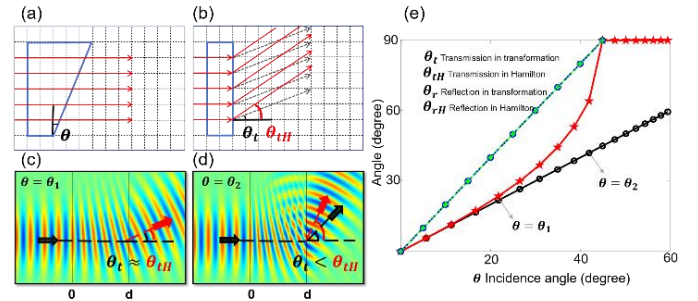

Fig. 1. Phase discontinuity in TO. (a) Virtual space, a vacuum trapezoid with a tilted angle $\theta$. (b) Physics space, a planar slab after compressed transformation. The phase is discontinuous at $x = d$, causing the transmission deviation between TO perspective and Hamiltonian optics. (c), (d) Simulated Gaussian wave patterns (Hz) in physical space of two different $\theta$ in the transformation. (e) The mathematical transmission and reflection of TO and Hamiltonian optics as $\theta$ changes.

---

Gaussian beam. Intriguingly, the reflection angle in the TO perspective and Hamiltonian optics are always the same ($\theta_r = \theta_{rH}$). It is because that the phase is uniform with continuous refractive index change at $x = 0$ [39]. To overcome the phase discontinuity and make the TO matches Hamiltonian optics, we put forward the MTO building upon the consistent reflections. Figure 2 elucidates the concept and basic principle of MTO. First, we put a perfect electric conductor (PEC) on the right side of the air trapezoid in virtual space and the incident parallel rays will be deflected, as shown in Fig. 2(a). Then, we use same compression transformation to transform the trapezoid into a planar slab but preserve the PEC wall (see Fig. 2(b)). The detailed transformation formula is as follows: $x' = \frac{(d+x)\cdot d}{d+\tan(\theta)\cdot x} - d$, $y' = y$, where we set $d = 2$ μm, $L = 8$ μm, and $\theta = \arctan(\frac{1}{4})$ and the permittivity and permeability in the slab are

$$\varepsilon' = \mu' = \begin{pmatrix} \frac{d^2 + \tan^2(\theta) \cdot (d+x')^2}{d(d+\tan(\theta) \cdot y')} & -\frac{\tan(\theta) \cdot (d+x')}{d} & 0 \\ -\frac{\tan(\theta) \cdot (d+x')}{d} & \frac{d+\tan(\theta) \cdot y'}{d} & 0 \\ 0 & 0 & \frac{d+\tan(\theta) \cdot y'}{d} \end{pmatrix} \quad (1)$$

The incident parallel light will still reflect, and importantly, the deflection angle of the rays leaving the slab is the same as in Fig. 2(a). This is because that the left boundary is keeping the same during the TO, and there is no phase discontinuity. Finally, we remove the PEC in Fig. 2(b) and add a mirror slab to from a composite structure, which we call it a MTO medium. Its dielectric tensor is satisfied with $\varepsilon'(x) = \mu'(x) = \varepsilon'(2d-x) = \mu'(2d-x)$. With the MTO medium, the reflected wave can be guided to the transmitted wave as depicted in Fig. 2(c), resulting in the emit angle twice than the tilted angle $\theta$ in physical space ($\theta_t = \theta_{tH} = 2\theta$). In Figs. 2(d)-(f), we choose Gaussian wave with transverse magnetic (TM) polarization for simulation. The $H_z$ fields are accorded with light rays in Figs. 2(a)-(c). We see that the right boundary of MTO medium eliminates the phase discontinuity and realizes the refractive index matching compared with Fig. 1(b), which evidence the feasibility of the MTO.

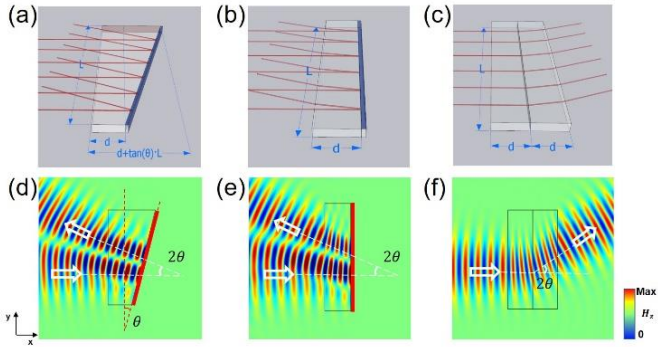

Fig. 2. The schematic of MTO. (a) Virtual space same as Fig. 1(a) but with PEC on the right side of the trapezoid. We set $d = 2$ μm, $L = 8$ μm, and $\theta = \arctan\left(\frac{1}{4}\right)$. The incident parallel rays are reflected with $2\theta$ (along the $x$ direction). (b) Physical space after compression transformation with PEC. The reflection behavior is the same as that of (a). (c) Physical space after MTO, owning a mirror symmetry of dielectric tensor with PEC, previous refractive rays are transformed into the transmitted rays with $2\theta$ (along the $x$ direction). Rays are calculated by Hamiltonian optics. (d)-(f) Corresponding simulation results (Hz field) of a TM Gaussian beam incidence of (a)-(c). The working wavelength $\lambda$ is set to be 1 μm.

In fact, the MTO method has addressed all the mismatches in optics. As the transformation induces $\varepsilon = \mu$, the impendence naturally matches. The phase discontinuity will be offset by introducing the symmetrical structure. The refractive index mismatches occur at boundaries where the refractive index changes dramatically. In the compression transformation, refractive index change is continuous at the incident boundary ($x = 0$). Hence, the symmetrical structure simultaneously resolves the refractive index mismatches. It is worth noting that the MTO makes good use of reflected wave and converted perfectly into transmission wave. This property is helpful to design the receiving devices.

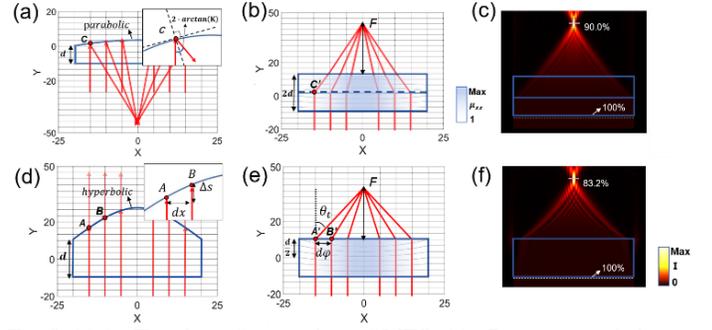

Fig. 3. (a)-(c) The planar focusing lens of MTO. (a) Rays in a parabolic PEC with $y_{para} = -\frac{x^2}{2h} + \frac{L^2}{8h}$ forms a imaging point at $(0, \frac{L^2}{8h} - \frac{h}{2})$. $L$=40μm, $h$=80 μm, d=10 μm. (b). Rays in the planar lens after MTO applying to (a). The background shows the gradient of $\mu_{xx}$. The imaging point is at (0, 37.5). (c) The simulated TM plane wave incident to the planar lens with a concentration ratio of 90.0%. (d)-(f) The planar focusing lens designed by generalized Snell law with $\frac{d}{2}$ shift in the y direction. The $\frac{d}{2}$ shift is introduced to facilitate better comparisons. (d) Rays in the virtual space, with the shape $y_{hyper} = -\sqrt{x^2 + F^2} + \sqrt{\frac{L^2}{4} + F^2}$. $F$=27.5 μm, d=20 μm. (e) Rays in physical space after compression transformation in (d). The imaging point is at $(0, F + \frac{d}{2})$. (e) The simulated TM wave of the planar lens with a concentration ratio of 83.2%. The working wavelength $\lambda$ is set to 1 μm.

Here we showcase a planar focusing lens of transmission types based on MTO. Simultaneously, we will compare this approach with another perfect focusing lens designed by generalized Snell law. They are of the same size and focal point. As shown in Fig. 3(a), a parabolic reflector (PEC) with a shape $y_{para} = -\frac{x^2}{2h} + \frac{L^2}{8h}$ will reflect the perpendicular incident parallel rays into a point imaging $(0, \frac{L^2}{8h} - \frac{h}{2})$ and vice versa. Such device is widely applied in antenna and solar energy collection. Using MTO ($y' = \frac{(d+y)d}{d+y_{para}} - d, x' = x$), this reflector can be transformed into a planar gradient lens in Fig. 3(b), with the parameters set as

$$\varepsilon'_{bottom} = \mu'_{bottom} = \begin{pmatrix} \frac{d+y_{para}}{d} & \frac{x' \cdot (d+y')}{h \cdot d} & 0 \\ \frac{x' \cdot (d+y')}{h \cdot d} & \frac{x'^2 \cdot (d+y')^2 + d^2 \cdot h^2}{d \cdot h^2 \cdot (d+y_{para})} & 0 \\ 0 & 0 & \frac{d+y_{para}}{d} \end{pmatrix} \quad (2)$$

at $y$<0; and

$$\varepsilon'_{top} = \mu'_{top} = \begin{pmatrix} \frac{d+y_{para}}{d} & \frac{x' \cdot (y'-d)}{h \cdot d} & 0 \\ \frac{x' \cdot (y'-d)}{h \cdot d} & \frac{x'^2 \cdot (y'-d)^2 + d^2 h^2}{d \cdot h^2 \cdot (d+y_{para})} & 0 \\ 0 & 0 & \frac{d+y_{para}}{d} \end{pmatrix} \quad (3)$$

at $y$>0. The parallel rays passed through the lens and imaging at the same distance as the reflection case. The simulation result of TM wave plane wave in Fig. 3(c) verifies the rays and reaches a 90% concentration ratio (the ratio of the average energy received by the absorber to the intensity of the incident energy). To demonstrate the superiority of the MTO approach, we construct another perfect lens based on the generalized Snell law. Suppose that gradient planar lens shown in Fig. 3(e) compressed by a vacuum area ($y' = \frac{(d+y)d}{d+y_{unknow}} - d, x' = x$) in Fig. 3(d) that can image at (0, F). The shape of the upper boundary needs to be solved. As analyzed before, there is phase discontinuity on the upper boundary of transformed planar lens. The transmission rays obey the generalized Snell law that

$$k_0 n_i \sin(\theta_i) dx + d\varphi - k_0 n_t \sin(\theta_t) = 0, \quad (4)$$

The wave vector $k_0 = 2\pi/\lambda$, $n_i$ and $n_t$ are refractive index of planar lens and outside medium, $\theta_i$ and $\theta_t$ are incident angle and transmission angle and $\theta_i = 0$, $n_t = 1$. The phase discontinuity $d\varphi$ can be calculated as

$$d\varphi = k_0 \sin(\theta_t) dx. \tag{5}$$

To make sure that all rays image at the same point (0, F), the transmission angle $\theta_t$ requires

$$\sin(\theta_t) = \frac{|x|}{\sqrt{x^2 + F^2}}. \tag{6}$$

Note that the phase discontinuity is original from the compression of virtual space that equals

$$d\varphi = k_0 dy. \tag{7}$$

From Eq. (5) and Eq. (7), we can figure the shape in Fig. 3(d) that

$$y_{unknow} = y_{\text{hyper}} = -\sqrt{x^2 + F^2} + \sqrt{\frac{L^2}{4} + F^2}. \tag{8}$$

Namely, the air area with the hyperbolic shape of Eq. (8) can be transformed into a planar focusing lens. These two lenses in Fig. 3(b) and Fig. 3(e) have the same focusing length. However, the hyperbolic transformed planar lens only provide with a 83.2% concentration ratio, which is much less than the MTO parabolic lens.

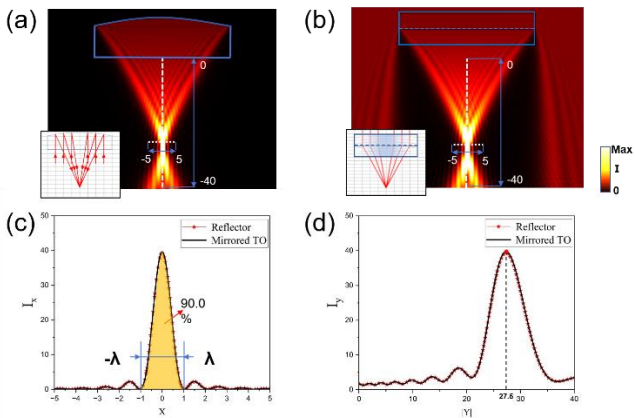

Fig. 4. The functionality comparison of MTO lens and ideal parabolic PEC (a)The reflection intensity field by a parabolic PEC. (b) The transmission intensity field (TM wave) through the focusing lens designed by MTO. The insets correspond to their ray trajectories and backgrounds show their permeabilities $\mu_{xx}$. (c) The horizontal intensity and (d) vertical distribution is matched closely comparing to the parabolic PEC with the MTO lens. The working wavelength λ is set to be 1 μm.

Finally, we emphasize that the lens designed by MTO has the same efficiency as the original parabolic PEC. As shown in Fig. 4(a), an upward incident plane wave will be refracted and focused at the imaging point. For comparison, the plane wave is incident on the MTO lens from top to bottom, which focuses on the same point. To be precise, the transverse and longitudinal intensity field profiles at the focal point are plotted in Figs. 4(c) and (d). It can be observed that the Hz intensity profiles are nearly identical. The concentration ratio of the parabolic reflector from -λ to λ (orange region in Fig. 4(c)) is the same with the lens designed by MTO. The significant overlap between the transmission and reflection fields confirms that the MTO allows for complete transmission and avoids material mismatch.

**Conclusion**

In brief, we propose a method called MTO to solve the material mismatch in transformation optics by introducing a mirrored medium to offset the phase discontinuities. MTO can control the direction of wave propagation just by transformation without calculation of generalized Snell's law or Hamiltonian optics. MTO makes up for the lack of shape-invariant transformation, and expands the scope of TO applications, which is of great value in many optical applications such as imaging and signal transmission.

**Funding.** The National Key Research and Development Program of China (2020YFA0710100); Jiangxi Provincial Natural Science Foundation (20224ACB201005); Fundamental Research Funds for the Central Universities (20720230102).